# DOUBLING THE INTENSITY OF AN ERL BASED LIGHT SOURCE *

Andrew Hutton, Jefferson Lab, Newport News, VA23693

*Abstract*

A light source based on an Energy Recovered Linac (ERL) [1] consists of a superconducting linac and a transfer line that includes wigglers and undulators to produce the synchrotron light. The transfer line brings the electrons bunches back to the beginning of the linac so that their energy can be recovered when they traverse the linac a second time, $\lambda/2$ out of phase. There is another interesting condition when the length of the transfer line is $(n\pm 1/4)\ \lambda$. In this case, the electrons drift through on the zero RF crossing, and make a further pass around the transfer line, effectively doubling the circulating current in the wigglers and undulators. On the third pass through the linac, they will be decelerated and their energy recovered. The longitudinal focusing at the zero crossing is a problem, but it can be canceled if the drifting beam sees a positive energy gradient for the first half of the linac and a negative gradient for the second half (or vice versa). This paper presents a proposal to use a double chicane at the center of the linac to provide this focusing inversion for the drifting beam while leaving the accelerating and decelerating beams on crest.

## INTRODUCTION

Light sources based on an Energy Recovered Linac (ERL) have been proposed at Cornell [2] and the 4GLS at Daresbury [3]. The ERL light source consists of a superconducting linac and a transfer line that includes wigglers and undulators to produce synchrotron light. The transfer line brings the electrons bunches back to the beginning of the linac so that their energy can be recovered when they traverse the linac. The usual requirement for electrons to recover energy when they transit the superconducting linac for the second time is that the length of the transfer line from the end of the linac back to the beginning of the linac should be $(n\pm 1/2)\ \lambda$.

There is another interesting condition when the length of the transfer line is $(n\pm 1/4)\ \lambda$. In this case, the electrons neither gain nor lose energy as they transit the linac the second time, drifting through on the zero RF crossing, and make a further pass around the transfer line, effectively doubling the circulating current in the wigglers and undulators. On the third pass through the linac, they will be decelerated as in the regular energy recovery mode and their energy will be recovered.

This proposal is a way to halve the current requirements on the gun for a given synchrotron light intensity. This is a significant reduction as this the most important technological challenge of ERL light sources. In addition, the higher order mode losses in the linac are reduced.

*Work supported by the U.S. DOE Contract No. DE-AC05-84ER40150

In a normal ERL, there are two passes through the linac (accelerating and decelerating), each with current I, so the HOM power is proportional to $2xI^2$. In this proposal, for the same synchrotron light intensity, there are a total of three passes through the linac (accelerating, drifting, and decelerating), each with current I/2, so the HOM power is proportional to $3x(I/2)^2 = 3I^2/4$. So the higher order mode losses in the linac are reduced by a factor 3/8 – another important advantage as this is probably the second most important technological challenge of ERL light sources.

However, there are problems with this simple scheme. On the second pass through the linac, bunches will be see a positive or negative longitudinal energy gradient depending on whether the path length is greater or less than an integer. During the third pass, the bunches are decelerated and there will tend to be beam loss at the lower energies as the effects of the longitudinal energy gradient combine with adiabatic anti-damping to blow up the emittance. In addition, users prefer all bunches to be identical as experiments would be hard to interpret with alternating bunch densities.

*Principle*

There is a way around this difficulty if the drifting beam sees a positive energy gradient for the first half of the linac and a negative gradient for the second half (or vice versa). Since the drifting beam has a different energy than the accelerating and decelerating beams at the linac center, the beams can be separated and their relative phases in the downstream linac independently optimized.

Let the injection energy into the linac be $E_I$ and the full energy of the beam at the end of the linac be $E_I + E_L$, where $E_I$ is much less than $E_L$ (for the Cornell ESR, $E_I$ = 10 MeV, $E_L$ = 6,900 MeV). At the midpoint of the linac, the energy during acceleration will be $E_I + E_L/2$, and the energy during deceleration will be the same at this point. However, the energy of the beam drifting through on the second pass will be $E_I + E_L$. If a chicane is installed at the mid point of the linac, the deflection angles of the magnets will be in the ratio $(E_I + E_L)/(E_I + E_L/2)$, or roughly, the accelerating and decelerating beams are deflected twice as much as the drifting beam. This can be used in one of two ways.

## METHOD 1 - $\lambda/2$ DELAY

If the chicane (in reality there would be two chicanes with common first and last magnets) is arranged to delay the accelerating/decelerating beams by $\lambda$ compared to the straight-ahead beam, the drifting beam will only be delayed by about $\lambda/2$. Careful design of the chicane can ensure that the difference in the delays between the two beams is exactly $\lambda/2$.

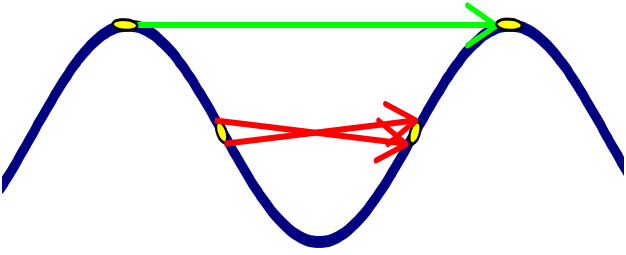

Figure 1: Principle of λ/2 Chicane

On the first pass, there is a λ phase delay in the chicane, which does not affect the acceleration. On the second pass, there is a λ/2 phase delay in the chicane, so that the beam sees a positive longitudinal gradient in the first half of the linac and a negative gradient in the second half. To first order, there is no longitudinal energy gradient at the end of the linac, and the beam is ready to transit the transfer line, wigglers and undulators. On the third pass, the beam is decelerated to $E_{inj}$ and can be deflected to a dump. But this simple proposal also has problems that make it of limited utility.

*Error analysis*

The energy spread in the beam is due to the injector energy spread (usually relatively small) and the bunch length that causes an energy spread on the first pass due to the curvature of the RF. The energy error $\delta E_{natural}$ at a distance $s$ from the center of a bunch that is on crest compared to the energy of the linac $E_L$ is given by

$$\delta E_{natural} / E_L = 1 - \cos(2\pi s/\lambda) \cong \pi^2 s^2/\lambda^2 \quad (1)$$

This sets the scale for comparing the size of the errors introduced by the chicane.

On the second pass through the linac, the head and tail of the drifting bunch will have slightly different energies at the chicane, and the delay induced by the chicane will, in general, be different for different energies. This can lead to a change in the bunch length and hence an increase in the energy spread of the bunch due to imperfect cancellation of the two halves of the linac.

Let us assume the simplest kind of chicane, with no quadrupoles and three short bends (θ, −2θ, θ), where θ is the deflection angle in radians of the beam of energy $E_c$ GeV. If the total straight-line length of the chicane is $L_c$, the difference in drift length $L_d$ for a particle of energy $E_c$ is equal to

$$L_d = L_c(1 - \cos\theta) \cong L_c\theta^2/2 \quad (2)$$

and

$$\theta = 0.3\, B_m L_m/E_c \quad (3)$$

where $B_m$ is the magnetic field in Tesla of the first chicane bend of length $L_m$. Combining equations 2 and 3 gives

$$L_d = L_c(0.3\, B_m L_m)^2/2E_c^2 \quad (4)$$

Differentiating to give the change in the path length difference as a function of energy error gives

$$\partial(L_d)/\partial E = -L_c(0.3\, B_m L_m)^2/E_c^3 = -2L_d/E_c \quad (5)$$

Let us now look at a particle in the drifting beam at a distance $s$ from the center of the bunch, which traverses the first half linac at the zero crossing (90° after crest). It will have an incoming energy error ΔE of

$$\Delta E = E_L/2 \cdot \sin(2\pi s/\lambda) \cong E_L \cdot \pi s/\lambda \quad (6)$$

The nominal energy for the chicane for the drifting beam is $E_c = E_L$ so the additional path length change is

$$\delta L = -(2L_d/E_L) \cdot E_L \cdot \pi s/\lambda = -2\pi s L_d/\lambda \quad (7)$$

The energy error $\delta E_{chicane}$ induced in the second linac by the additional chicane delay can be found by combining equations 5 and 7

$$\delta E_{chicane}/E_L = -2\sin(2\pi\delta L/\lambda) = -2\pi^2 s L_d/\lambda^2 \quad (8)$$

This should be compared to the energy spread, $\delta E_{natural}$, produced by the curvature of the RF in the first pass acceleration of the beam.

$$\delta E_{chicane}/\delta E_{natural} = -2\pi^2 s L_d/\lambda^2 / \pi^2 s^2/\lambda^2 = -2L_d/s \quad (9)$$

The design delay of the chicane $L_d$ is equal to λ/2 so

$$\delta E_{chicane}/\delta E_{natural} = -\lambda/s \quad (10)$$

This error is sufficiently large that it must be addressed. There are two alternatives that will be examined separately: reducing the $M_{56}$ in the chicane, and inverting the bunch head and tail.

*Reducing $M_{56}$*

The problem with the simple chicane shown in Figure 1 is that the delay is proportional to the energy error so the cancellation of the two linacs is destroyed. If a chicane is designed such that the path length is independent of the energy error, the problem is solved. Designs for such a chicane always require more space than the simple three-bend chicane discussed above. The requirement can be stated in terms of the $M_{56}$ matrix element (also known as $R_{56}$) as

$$M_{56} = \partial(L_d)/\partial E = \int \frac{D_x}{\rho}dz = 0 \quad (11)$$

So the horizontal dispersion $D_x$ must be managed in the bends where the bending radius ρ is non-zero, which implies the use of additional quadrupoles and bends to meet the matching conditions required. This condition can be met under specific conditions, such as the proposed CEBAF proof of principle experiment [4], but in general is not a convenient design choice

There is, however, another solution (suggested by Mike Tiefenback).

## METHOD 2 - HEAD-TAIL INVERSION

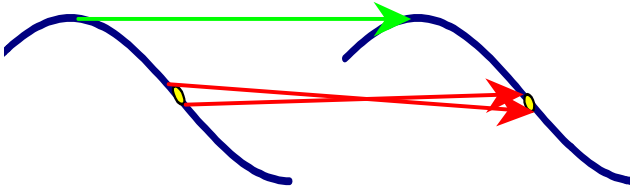

Figure 2: Principle of head-tail inversion

The idea is that the chicane has just the right delay so that the head and tail exchange positions as shown schematically in Figure 2. This will occur if the change in position of a particle at distance **s** from the bunch center δL is equal to –2**s**. Then from Equation 7

$$\delta L = -2\mathbf{s} = -2\pi \mathbf{s} L_d / \lambda \qquad (12)$$

$$L_d = \lambda/\pi \qquad (13)$$

This result gives a simple, exact criterion for designing the drift chicane. The $M_{56}$ matrix element should be

$$M_{56} = \partial(L_d)/\partial E = -\lambda/E_L$$

## CHICANE DESIGN

### The drift chicane

The principal parameters are as follows: the linac energy is $E_L$ and the drift chicane delay is $L_d = \lambda/\pi$. The chicane can be a simple design with no quadrupoles and three bends (θ, –2θ, θ). To illustrate the design, the parameters of the Cornell ERL will be used: $E_L$ = 7 GeV, λ = 0.23077 meters (1300 MHz), so from Equation 13, $L_d$ = 0.0732 meters.

The difference between the path length in the magnet and the length of the magnet, $L_m$, is given by

$$\Delta L_m = L_{magnet}(\theta/\sin\theta - 1) \qquad (14)$$

while the difference between the trajectory and the drift length $L_D$ is given by

$$\Delta L_D = L_{drift}(1/\cos\theta - 1) \qquad (15)$$

So the path difference created by a four bend chicane with two equal drift lengths between magnet 1&2 and 3&4 is

$$L_d = 4L_{magnet}(\theta/\sin\theta - 1) + 2L_{drift}(1/\cos\theta - 1) \qquad (16)$$

For a practical magnet, $L_{magnet}$ ~ 2 meter and $B_m$ ~ 1.75 Tesla, so the bend angle θ ~ 0.15 radian for an energy of 7 GeV. If the drift length is also 2 meters, the path length difference given by Equation 16 is 0.0755 meters, i.e. roughly the right amount. The total length of this chicane is about 14 meters, similar to the length of the cryomodule (9 meters).

### The Double Chicane

It is difficult to come up with a design for the low energy beam chicane in which the delay is the same as the drift chicane. Because the energy is only half, the natural delay using the same initial bends would be roughly four times as long. A better solution is to add an additional wavelength delay so $\Delta L_C = \lambda + \lambda/\pi$. The desired ratio of the delays for the two chicanes is then π + 1 = 4.142, about what would be obtained naturally. Using the same magnets and spacing as the case considered above, $L_m$ ~ 2 meter, $B_m$ ~ 1.75 Tesla, and for an energy of 3.5 GeV, θ ~ 0.30 radian, and if the drift length is also 2 meters, the path length difference is 0.313 meters, 4.083 times the delay in the drift chicane, i.e. roughly the right amount. Getting the delay exactly right can be done by small variations of the drift lengths between magnets 1&2 and 3&4. The two chicanes therefore look like Figure 3.

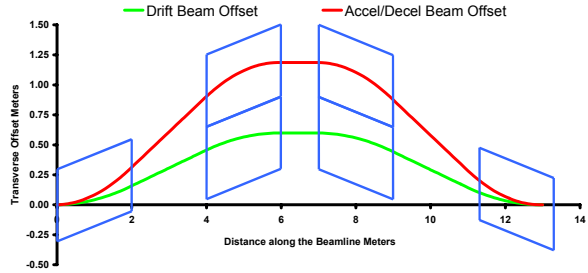

Figure 3 Double Chicane Design

## SET-UP

The linac phases are crested for maximum $E_L$ with the chicane off. The chicane is turned on and the magnet excitation tuned to give exactly the same $E_L$ as the single pass beam, at which point the chicane is set correctly for the accelerating beam. The beam is then brought around the transfer line and drifted down the first half of the linac. The delay of the transfer line is adjusted to provide the correct energy at the midpoint of the linac, corresponding to the zero crossing of the RF in the first half of the linac, by using the first chicane magnet as a spectrometer. The delay of the high-energy chicane is then adjusted to minimize the bunch length at the end of the linac, corresponding to the other zero crossing in the second half of the linac. On the third pass, the beam is decelerated down the first half of the linac, receiving a λ delay at the chicane and being decelerated in the second half of the linac.

## REFERENCES
[1] G. R. Neil, et al, Phys. Rev. Let. 84, 662 2000
[2] Eds. S.M. Gruner, M. Tigner, CHESS Technical Memorandum 01-003, aka JLAB-ACT-0104, 2001
[3] M W Poole et al, "4GLS: an Advanced Multi-source Low Energy Photon Facility for the UK", EPAC'02, Paris.
[4] Jefferson Lab Experiment E02-102, A. Bogacz, A. Hutton, Spokespersons